# Large exchange-dominated domain wall velocities in antiferromagnetically coupled nanowires

Majd Kuteifan*[1], M. V. Lubarda[2], S. Fu[1], R. Chang[1], M. A. Escobar[1], S. Mangin[3], E. E. Fullerton[1], and V. Lomakin[1]

[1]University of California San Diego, San Diego, United States of America;

[2]Faculty of Polytechnics, University of Donja Gorica, Podgorica, Montenegro;

[3]Institut Jean Lamour UNR 7198 CNRS Universite de Lorraine, Vandoeuvre- Les-Nancy, France;

*email : makuteif@eng.ucsd.edu

**Abstract**

Magnetic nanowires supporting field- and current-driven domain wall motion are envisioned for new methods of information storage and processing. A major obstacle for their practical use is the domain-wall velocity, which is traditionally limited due to the Walker breakdown occurring when the forcing field or current reaches a critical threshold value. We show through numerical and analytical modeling that the Walker breakdown limit can be extended or completely eliminated in antiferromagnetically coupled magnetic nanowires. These coupled nanowires allow for giant domain-wall velocities driven by field and/or current via spin transfer torque as compared to conventional nanowires.

Manipulating magnetic domain walls (DWs) to store and transfer information is envisioned to enable high-density, low-power, non-volatile, and non-mechanical memory, recording and processing systems. Related concepts have been explored in the past, e.g. bubble memory [1], and are promising for future systems, e.g. racetrack memory [2] where DWs can be moved by applied magnetic fields [3] and/or by currents [4] via the spin transfer torque (STT) effects [5, 6]. STT arises from the transfer of angular momentum from spin-polarized electrons to the DW magnetic moments and provides particularly attractive opportunities for DW manipulation [7]. However, there are several obstacles to be overcome to enable these technologies. One obstacle is the Walker breakdown limit [8, 9], which imposes a maximum velocity on the DW motion in magnetic systems posing a major problem in terms of the information transfer and storage speed. The Walker breakdown limit originates from the demagnetization field that imparts to the magnetization a torque which takes the opposite direction once the applied field and/or current exceed a certain critical value. Below these critical field or current values the DW velocity increases with increasing field or current, whereas above this critical value the DW motion exhibits back and forth oscillations. If the Walker breakdown effect could be eliminated then the increased DW speed would allow a major improvement in terms of data rates. Approaches have been presented for reducing or eliminating the Walker breakdown, such as those based on complex topologies or on alternative physical effects [10-12]. However, these approaches may be hard to implement in practical systems. Moreover, no approach has been shown to eliminate the Walker breakdown for DW motion induced by STT effects.

In this letter we present both an analytical model and numerical simulations of antiferromagnetically coupled nanowires (AFC NWs) subject to applied fields and currents. We find that this structure extends or completely eliminates the Walker breakdown limit for DW motion induced by fields or STT currents. For field-driven DW motion the maximum DW velocities and corresponding applied fields can be much higher than those of single-phase NWs. For STT induced DW motion the Walker breakdown can not only be reduced but also eliminated. Importantly, this structure can be readily realized experimentally.

The proposed structure is made of two antiferromagnetically coupled magnetic layers as shown in Fig. 1. The specific model used is made of two stacked soft magnetic NWs antiferromagnetically coupled through their common interface. The saturation magnetization of the first (top) and second (bottom) NWs are $M_{s1}$ and $M_{s2}$, respectively. The antiferromagnetic coupling between the NWs is sufficiently strong such that a single DW across both NWs is present (Fig. 1).

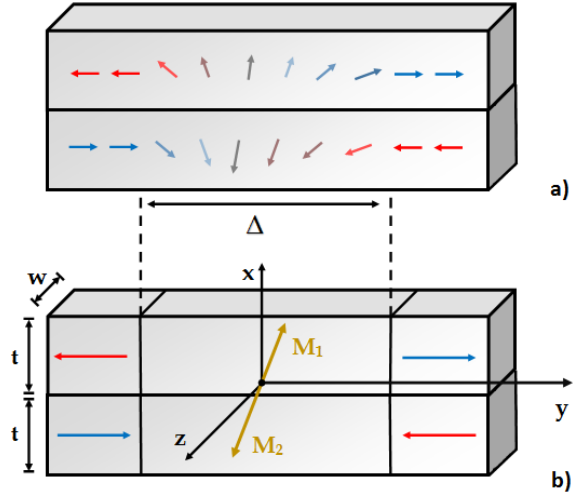

Figure 1: Schematic of a DW in an AFC NW. (a) The DW represented by a continuous variation of the magnetization states; (b) Representation in the analytical model in which the DW in each layer is represented by a single macrospin with its width $w$, thickness $t$, and length $\Delta$.

The dynamics of a DW in a NW under the influence of field and/or current are governed by the Landau-Lifshitz-Gilbert equation extended to include the STT effect [7] [9]:

$$\frac{\partial \mathbf{M}}{\partial t} = \gamma \mathbf{H}_{\text{eff}} \times \mathbf{M} + \frac{\alpha}{M_s} \mathbf{M} \times \frac{\partial \mathbf{M}}{\partial t} - u \frac{\partial \mathbf{M}}{\partial y} + \frac{\beta u}{M_s} \mathbf{M} \times \frac{\partial \mathbf{M}}{\partial y} \qquad (1)$$

In the above equation, $\mathbf{M}$ is the magnetization, $\gamma$ is the gyromagnetic ratio, $\mathbf{H}_{\text{eff}}$ is the effective field, $\alpha$ is the damping parameter, $M_s$ is the saturation magnetization, $\beta$ is the non-adiabatic spin-transfer paramater, and the parameter $u$ depends on the current density $J$ and is defined as $u = gJ\mu_B P/(2eM_s)$, where $g$ is the Landé factor, $\mu_B$ is the Bohr magneton, $e$ is the electron charge, and $P$ is the polarization factor of the current.

The DW dynamics in a single-phase NW governed by this equation can fall into different regimes depending on the values of the described parameters. In the absence of current, if the external field intensity is below the Walker threshold, the wall moves at a constant velocity [13]. If the applied field is stronger than this critical field, there is precession, intervals of backward motion and overall slowdown of the DW propagation. When only current is applied, the dynamics depend on the relative values of the damping parameter $\alpha$ and the non-adiabatic spin-transfer parameter $\beta$. If $\beta = \alpha$, the DW motion is steady for any DC applied current. If $\beta > \alpha$, the DW motion is steady for currents smaller than a certain limit, but slows down for stronger currents due to the onset of precession and backward motion. If $\beta < \alpha$, there is a range of low currents for which the spins of the DW tilt out of plane, after which the DW is stationary, i.e., no motion. For yet stronger currents, the DW propagates during which its magnetization undergoes precession. All these limits correspond to the Walker threshold that depends on the damping parameter $\alpha$, the saturation magnetization $M_s$ and the geometry of the NW.

In an AFC NW system it is possible to alter the symmetry of the problem so as to use exchange fields to compensate torque terms that lead to DW instability. This AFC NW geometry significantly reduces the Walker breakdown effects and even eliminates the Walker breakdown when current is used to move the DWs via the STT effect. This results in a drastic increase of the DW velocities in a simple geometry that is practically feasible using different materials such as in-plane anisotropy CoFeB/Ru/CoFeB or Fe/Cr/Fe structures or out-of-plane [Co/Ni]/Ru/[Co/Ni] structures.

The presented simulation results were obtained by solving the LLG equation using both FastMag and OOMMF simulators, which gave nearly identical results. These simulators respectively use finite element method (FEM) and finite differences method and provide a variety of tools that can be used to study DW motion. For each simulation the DW was located inside the NW and its position was monitored. The DW speed in the numerical simulations was obtained by calculating the time for the DW to propagate over a fixed distance (chosen as 10 micron). We studied the evolution of the DW velocity as a function of the applied field in single-phase and AFC NW for different values of saturation magnetization, damping and exchange fields. For the AFC NW case, each NW had a 4 nm thickness and a 20 nm width. The single-phase NW had an 8 nm thickness and 20nm width.

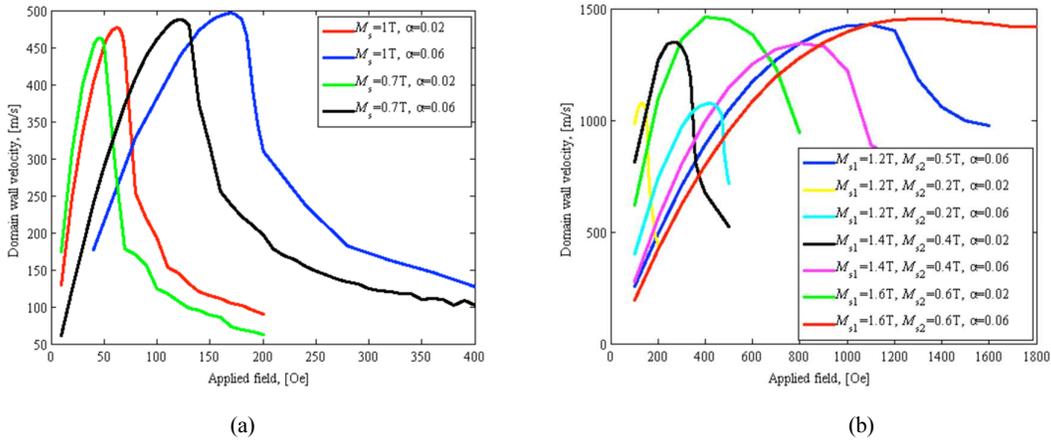

(a)     (b)

Figure 2: Average domain velocities driven for applied magnetic field obtained via FEM micromagnetic simulations. (a) single-phase NW with different values of the damping and saturation magnetization; ( b) AFC NW with different values of the damping and saturation magnetization with exchange energy density of 50 erg/cm.

Figures 2a and b show the DW velocity versus the applied field for single-phase and AFC NW, respectively. The results are given for different damping constants and saturation magnetization values. The magnetocrystalline anisotropy was kept to zero in both models, hence the anisotropy was entirely due to magnetostatics (shape anisotropy). However, this approach is also applicable to perpendicular anisotropy systems. For a single-phase NW, changing the saturation magnetization and damping only affects the DW mobility and the Walker breakdown critical field, but not the maximum achievable DW velocity in the system, as seen from Fig. 2a. This limitation clearly does not hold for AFC NW (Fig. 2b), where the mobility, Walker breakdown critical field, and peak velocity can be modulated through the saturation magnetization and damping of the constituent layers (Fig. 2b). Indeed, even when the total saturation magnetization of the composite system $|M_{s1} - M_{s2}|$ is identical to the saturation magnetization of the single-phase NW, the DW motion characteristics are different (Figs. 2a, b). The interaction between the two antiferromagnetically coupled layers and the symmetry of the system under the applied field must be taken into account to explain this phenomenon.

Several observations from Fig. 2b can be made. First, the closer the saturation magnetizations $M_{s1}$ and $M_{s2}$ are to each other the greater the peak velocity. However, in the pre-Walker breakdown regime, the smaller the net magnetization the slower the DW motion for a given magnetic field. The mobility, defined as the rate of change of DW velocity with the applied field $dv/dH$, is therefore proportional to the net magnetization of the AFC NW. If the saturation magnetization is equal in both layers, there is no motion under an applied field.

The operation of the AFC NW can be understood by considering the compensation of torques in the AF-coupled system, which is mediated through antiferromagnetic interlayer exchange. On account of the relative orientations of the Zeeman and demagnetization fields at each layer for the coupled system of moments, the precessional Zeeman torques compensate, while the precessional demagnetization and damping torques complement each other As a consequence of the altered dependence of the torque magnitudes on saturation magnetization, the Walker breakdown occurs for much greater fields and far greater DW velocities. However, increases in peak velocities are accompanied with a reduced DW mobility when the DW is moved by field.

The situation is different when the DW is moved by current. The difference is due to the fact that the symmetry of the system changes and the characteristics describing motion for the same parameter sets are significantly different.

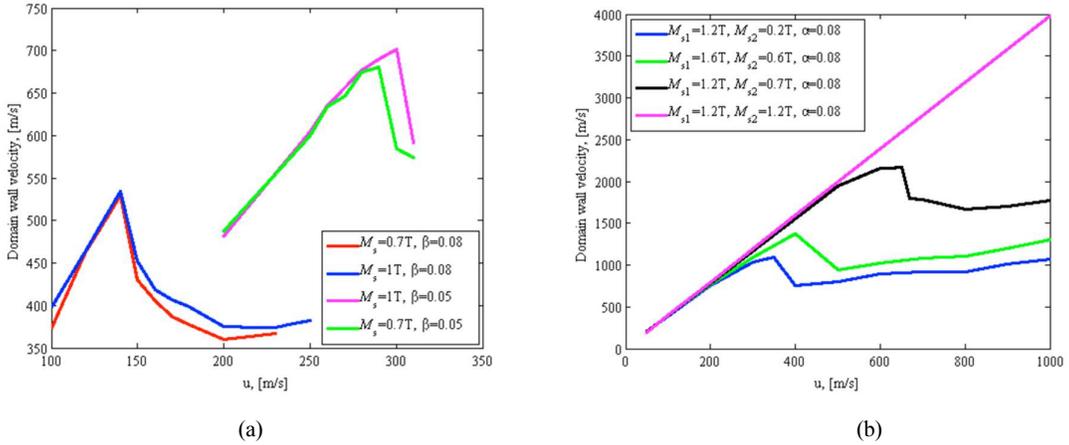

Figure 3: DW velocity as a function of u obtained via FEM micromagnetic simulations. (a) single-phase NW with different values of the non-adiabatic parameter and saturation magnetization; (b) AFC NW with different values of the non-adiabatic parameter and saturation magnetization with exchange energy density of 50 erg/cm.

Figure 3 shows the velocity of a DW in AFC NWs for different values of the applied current, the non-adiabatic parameter β and $M_{s2}$. In the case where $M_{s1} = M_{s2}$ the Walker breakdown is not encountered for any value of current amplitude for the AFC NW. This is because under current bias, the adiabatic spin-transfer torques of both layers compensate, preventing the spins to tilt out of plane and to trigger Walker breakdown. When $M_{s1}$ and $M_{s2}$ are different the Walker breakdown occurs at lower currents when β is large compared to α but this criterion also makes the steady state propagation of the DW faster.

For both field and current bias, the simulations indicate that the peak velocities and the Walker breakdown threshold are higher for an AFC NW than for single-phase NW. In both cases, the outcome is due to the compensation of different torque terms on account of system symmetry under given bias, which is mediated though the interlayer exchange interaction.

To gain a better understanding of this phenomenon, we compared the results to an analytical model. The motion of the DW is due to the torques applied on the magnetic moment. By calculating all these torques, it is possible to find the DW velocity and the Walker breakdown limit. This method has been described in detail for a conventional single-phase NW by A. Mougin *et al.* in ref [15]. It is equally applicable to systems with in-plane and out-of-plane magnetic anisotropies.

In this model, the DWs are considered as two macrospins with saturation magnetizations $M_{s1}$ and $M_{s2}$ (Fig. 1b). The antiferromagnetic coupling is considered to be infinite, such that the two macrospins are oriented perfectly opposite to one another. Calculating all the torques applied on the macrospins using spherical coordinates, we can link them to the movement of the DW by using the angular momentum conservation $d\mathbf{L}_{total}/dt = \Gamma_{total}$. The obtained net precessional torques due to the Zeeman and demagnetization field, and the net damping torque for the AFC case, relative to the corresponding torques for the single-phase NW, are $\Gamma_H^{AFC} = (M_{s1} - M_{s2})\Gamma_H$, $\Gamma_d^{AFC} = (M_{s1}^2 + M_{s2}^2)\Gamma_d$, and $\Gamma_\alpha^{AFC} = (M_{s1} + M_{s2})\Gamma_\alpha$. Solving the angular momentum conservation equation for the case of steady DW motion (i.e., motion during which the DW structure does not change) leads to the Walker threshold condition for an AFC NW:

$$H_W^{AFC} + \frac{M_{s1} + M_{s2}}{M_{s1} - M_{s2}} \frac{u}{\gamma\Delta}(\beta - \alpha) = 2\pi\alpha \left| N_y - N_x \right| \frac{(M_{s1}^2 + M_{s2}^2)(M_{s1} + M_{s2})}{(M_{s1} - M_{s2})^2}, \quad (1)$$

where $N_x$ and $N_y$ are the demagnetizing factors of the volume containing the DW, $\Delta$ is the length of the DW, and $H_W^{AFC}$ is the Walker field for the AFC case. For the single-phase NW case, this reduces to the following condition [16]

$$H_W^{single} + \frac{u}{\gamma\Delta}(\beta - \alpha) = 2\pi\alpha \left| N_y - N_x \right| M_{s1}, \quad (1)$$

where $H_W^{single}$ is the Walker field for the single-phase NW.

It is also possible to find the DW velocity in AFC NW for any value of field and current below the Walker breakdown limit:

$$v = \frac{\Delta\gamma}{\alpha}\left(H \frac{M_{s1} - M_{s2}}{M_{s1} + M_{s2}} + \frac{\beta}{\gamma\Delta}u\right), \quad (1)$$

. The result for the single-phase NW case is

$$v = \frac{\Delta\gamma}{\alpha}\left(H + \frac{\beta}{\gamma\Delta}u\right), \quad (1)$$

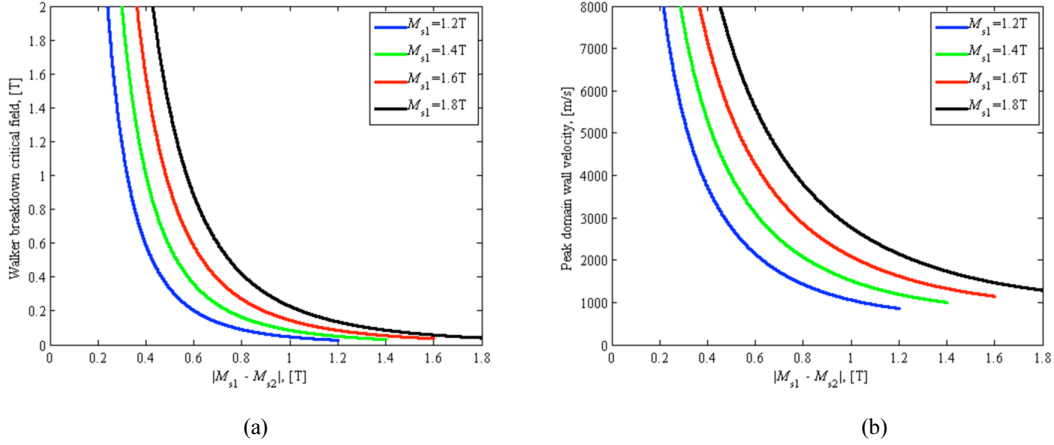

Figure 4: (a) Walker breakdown critical field vs. $|M_{s1} - M_{s2}|$ for AFC NWs for different values of $M_{s1}$; (b) Peak DW velocity at this critical field vs. $|M_{s1} - M_{s2}|$ for AFC NWs for different values of $M_{s1}$. The results were obtained via the analytical formulation for AFC NWs with a damping constant of 0.06. The FEM micromagnetic results were within 10% of the analytical results.

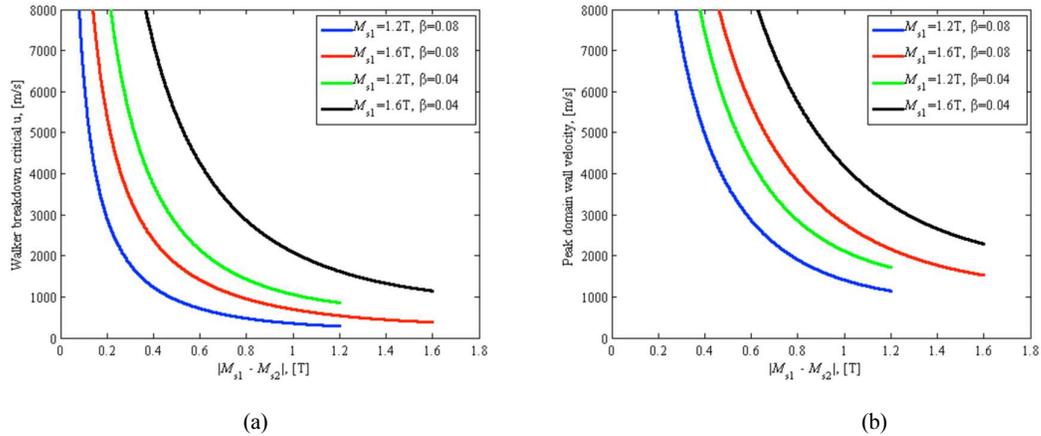

Figure 5: (a) Walker breakdown critical $u$ vs. $|M_{s1} - M_{s2}|$ for AFC NWs for different values of $M_{s1}$ and $\beta$; (b) Peak DW velocity at this critical field vs. $|M_{s1} - M_{s2}|$ for AFC NWs for different values of $M_{s1}$ and $\beta$. The results were obtained via the analytical formulation for AFC NWs with a damping constant of 0.02. The FEM micromagnetic results were within 10% of the analytical results.

Figures 4 and 5 show the Walker threshold field and current, respectively, and the corresponding maximal DW velocities as a function of the saturation magnetization difference $|M_{s1} - M_{s2}|$. It is evident that the Walker threshold strongly depends on the respective values of $M_{s1}$ and $M_{s2}$. As $|M_{s1} - M_{s2}|$ decreases, the Walker breakdown limit is progressively differred, and in principle can be made as large as needed or as is allowed by other limitatons, such as heat or disorder. The downside is that making the saturation magnetization of both layers very close dramatically reduces the mobility of the DW for the case of field driven DW propagation. However, for the case where only current is applied to the NW, the DW velocity does not depend on the saturation magnetizations and the equation becomes $v = u\beta/\alpha$ the same as for a single-phase NW in the pre-Walker breakdown regime. Therefore, using close or identical values for $M_{s1}$ and $M_{s2}$, it is possible to eliminate the Walker breakdown in current operated AFC NWs without imparing the mobility of the DW, thus achieving very high DW velocities.

In conclusion, we have shown that the maximum achievable DW velocity in AFC NWs under field or current bias can far exceed velocities attainable using single-phase NW systems. This result is not a consequence of a reduced average moment of the AFC-NW. We attributed this effect to the particular symmetry of the described system under a given bias, and the consequent compensation of torque terms that are responsible for structural instability. It was demonstrated that the Walker breakdown field could be significantly deferred by choosing saturation magnetizations of the two layers to be comparable. The characteristics of DW motion under field and current bias where demonstrated using micromagnetic simulations, and further investigated by an analytical model. The results obtained by the model were in good agreement with the simulation results, and produced formulas predicting the Walker breakdown field, peak velocity under field and current bias, and wall mobility in AFC systems. It was shown that it is possible to suppress the Walker breakdown limit for current-driven DW motion without impairing its mobility by using similar values of the saturation magnetization in both layers of the AFC NW. This can be done for any value of the damping parameter α or the non-adiabatic spin-transfer torque β. Moreover the AFC structure makes this type of NW weakly sensitive to residual magnetic fields thanks to the lowered total magnetization. The proposed system and described phenomena are expected to be interesting for future theoretical and experimental investigations of DW dynamics, as well as for technological applications related to data storage and processing, such as race track memories.

**Acknowledgments**

This work was supported by the NSF grants DMR-1312750 and CCF-1117911, ANR-13-IS04-0008-01 " COMAG" and Partner University Fund 'Novel Magnetic Materials for Spin Torque Physics', and European Project (OP2M FP7-IOF-2011-298060) and the Region Lorraine.